\documentclass[conference, 10pt]{IEEEtran}
\IEEEoverridecommandlockouts
\usepackage{cite}
\usepackage{amsmath,amssymb,amsfonts}
\usepackage{algorithmic}
\usepackage{graphicx}
\usepackage{textcomp}
\def\BibTeX{{\rm B\kern-.05em{\sc i\kern-.025em b}\kern-.08em
    T\kern-.1667em\lower.7ex\hbox{E}\kern-.125emX}}

\usepackage{csquotes}
\usepackage[hyphens]{url}

\usepackage{caption}
\usepackage{subcaption}
\usepackage{dblfloatfix}
\usepackage[inline]{enumitem}
\usepackage{tabularx}
\usepackage{booktabs}
\usepackage{pifont}
\usepackage[dvipsnames]{xcolor}

\newcommand{\cmark}{\textcolor{Green}{\ding{51}}}
\newcommand{\xmark}{\textcolor{Red}{\ding{55}}}

\usepackage{hyperref}

\begin{document}

\title{Sustainable Cooperation in Peer-To-Peer Networks
}

\author{
 \IEEEauthorblockN{Bulat Nasrulin, Rowdy Chotkan, Johan Pouwelse}
 \IEEEauthorblockA{Delft University of Technology
 \\\{b.nasrulin, r.m.chotkan-1, j.a.pouwelse\}@tudelft.nl}
}

\maketitle
\thispagestyle{plain}
\pagestyle{plain}

\begin{abstract}
Traditionally, peer-to-peer systems have relied on altruism and reciprocity. Although incentive-based models have gained prominence in new-generation peer-to-peer systems, it is essential to recognize the continued importance of cooperative principles in achieving performance, fairness, and correctness. The lack of this acknowledgment has paved the way for selfish peers to gain unfair advantages in these systems. As such, we address the challenge of selfish peers by devising a mechanism to reward sustained cooperation. Instead of relying on global accountability mechanisms, we propose a protocol that naturally aggregates local evaluations of cooperation. Traditional mechanisms are often vulnerable to Sybil and misreporting attacks. However, our approach overcomes these issues by limiting the benefits selfish peers can gain without incurring any cost. The viability of our algorithm is proven with a deployment to 27,259 Internet users and a realistic simulation of a blockchain gossip protocol. We show that our protocol sustains cooperation even in the presence of a majority of selfish peers while incurring only negligible overhead.
\end{abstract}


\section{Introduction}

Despite a wealth of experimentation with incentives in peer-to-peer networks, even such well-developed projects as Bitcoin and BitTorrent~\cite{bittorent} do not directly address the fundamental problem of cooperation in these systems. The profit-seeking incentive for Bitcoin miners on its own does not provide guarantees for the sharing of historical blocks, which is accomplished on an altruistic basis~\cite{sapirshtein2017optimal}. BitTorrent, as well, critically relies on pure altruism for content seeding, after the tit-for-tat phase~\cite{piatek2007incentives}.

As such, selfish nodes pose a significant risk to the overall health and functioning of peer-to-peer networks. More specifically, the Bitcoin blockchain is vulnerable to a type of attack known as \textit{selfish mining}~\cite{selfishmining2018}. In this deceptive mining strategy, nodes withhold successfully mined blocks to create a fork and get ahead of the longest public chain.
This leads to two issues: first, it creates a scenario in which dishonest nodes have an unfair advantage and receive more mining rewards, and second, it degrades network performance~\cite{selfish_mining_impacts_2021}. Both of these problems have detrimental effects. The former issue results in a situation where honest nodes either stop participating altogether due to a lower probability of receiving block rewards or adopt the selfish mining strategy to increase their personal gain. Both of these scenarios, in turn, lead to more centralization within the Bitcoin network, further exacerbating the centralization imposed by mining pools~\cite{bitcoin_central2015}. Meanwhile, the latter issue leads to higher delivery times for honest blocks and opens up the possibility of attacks such as double spending due to forks appearing in the network~\cite{selfish_mining_2021}.

The BitTorrent protocol~\cite{bittorent}, introduced over two decades ago, suffers from a similar issue. As mentioned, in this protocol, the availability of files is also dependent on altruism and reciprocity exhibited by nodes~\cite{piatek2007incentives}. However, similarly, there is no direct deterrent against nodes that do not contribute to the availability of files in the network. This behavior was referred to as \textit{free-riding}, characterizing nodes that only receive files and do not aid in sharing them with the network~\cite{freeriding2004}. Exhibiting this behavior is not an optimal strategy for nodes, as it decreases the overall availability of files and thus the health of the network.

\begin{figure}[t]
    \centering
    \begin{subfigure}[b]{\linewidth}
        \centering
        \includegraphics[width=.65\linewidth]{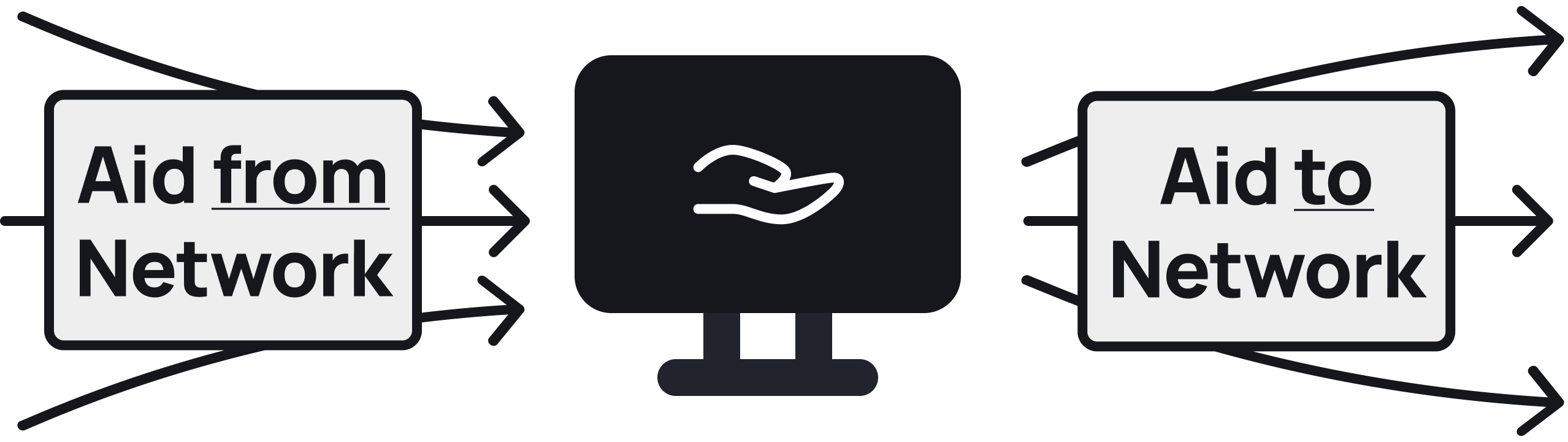}
        \caption{Honest}    
        \label{fig:selfish_a}
    \end{subfigure}
    \hfill
    \vspace{.1em}
    \begin{subfigure}[b]{\linewidth}  
        \centering 
    \includegraphics[width=.65\linewidth]{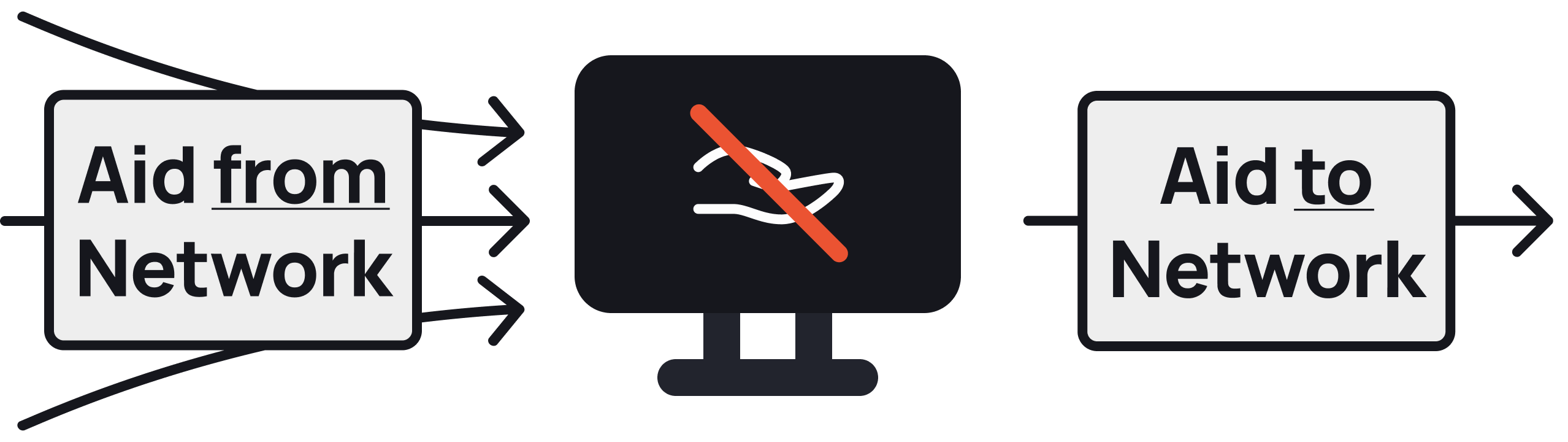}
        \caption{Selfish}    
        \label{fig:selfish_b}
    \end{subfigure}
    \caption{The behavior of nodes.}
    \label{fig:selfish}
\end{figure}

We argue that both of these examples are specific instances of a more general problem that we label as \textit{selfish behavior}. It refers to any type of behavior exhibited by nodes in peer-to-peer systems where they withhold resources (data) for personal gains even when it is detrimental to the overall health of the system. In the case of Bitcoin, this manifests itself as the withholding of blocks or transactions, while in BitTorrent, it manifests as \textit{leeching}: downloading but not sharing files with others.~\autoref{fig:selfish} visualizes this concept: honest nodes aid the network by sharing information (e.g., transactions or files), whereas selfish nodes merely reap the benefit of the services provided to them by the network while either sharing nothing or sharing selectively with colluding nodes.

Furthermore, we argue that while this problem is prominent in the Bitcoin and BitTorrent protocols, it is not unique to them. Rather, it is a fundamental problem that all peer-to-peer systems are susceptible to. This problem is rooted in the absence of efficient mechanisms for the accountability of peers at the networking layer: in both instances, a lack of sharing information or data does not directly lead to negative consequences for the offending peer, though they may in the long term as they essentially poison the network by deteriorating its performance (e.g., the availability of files).

Proposed solutions (e.g., see PeerReview~\cite{peerreview2007} and LiFTinG~\cite{lifting2010}) show promising results with respect to identifying malicious nodes. However, they introduce hefty overheads by relying on the analysis of all outgoing and incoming messages, the requirement of interactions with all nodes within a network, or by being too tailored to be applicable to any type of peer-to-peer system. There is also another largely unaddressed issue prominent in all permissionless networks: the Sybil attack. Enabled by the fact that it is trivial for selfish peers to re-enter the collaborative network under a new identity on multiple instances after abuse of a system ~\cite{douceur2002sybil}. Without active consideration of this attack, any solution becomes vulnerable to manipulation. As such, there is a need for an accountability mechanism that can: 1) exclude nodes that consistently abuse the network, addressing the risk imposed by Sybils; and 2) functions in any peer-to-peer system without introducing unnecessary overhead or complexity.

To address these needs, we propose an accountability system that is based on an evolutionary mechanism referred to as \textit{indirect reciprocity}~\cite{nowak2005evolution,schmid2021unified}: a more advanced mechanism rather than the \textit{tit-for-tat} strategy~\cite{cohen2003incentives}. Our mechanism takes indirect contributions into account when assessing the trustworthiness of nodes. This mechanism also does not rely on global trust scores but accounts for the trustworthiness of peers with their local connections. We argue that \emph{locality} is both necessary and sufficient in achieving sustained cooperation for peer-to-peer networks. In order to mitigate \emph{selfish behavior} and promote fair resource allocation, our approach takes the \emph{subjective contributions} of nodes into account, allowing us to overcome issues of misreporting and manipulation by Sybils. Furthermore, we make use of the MeritRank algorithm~\cite{nasrulin2022meritrank} to assign subjective trust scores to nodes.

To summarize, our work makes the following contributions:
\begin{enumerate}
\item We present a generic accountability mechanism based on indirect reciprocity, which is able to function in any type of peer-to-peer network.
\item We evaluate this mechanism by analyzing its performance through a simulation of the Bitcoin network and a deployment on the network of a BitTorrent client~\cite{tribler}, showcasing how our mechanism rewards well-performing peers based on subjective contributions and handles reputations based on indirect reciprocity.
\end{enumerate}

The remainder of this paper is structured as follows: ~\autoref{sec:system_model} describes the system model in which we operate and formulates the problem we are solving. Next, ~\autoref{sec:solution} discusses our design and in~\autoref{sec:results} we perform a performance analysis of our proposed mechanism. Finally,~\autoref{sec:related_work} discusses related work and~\autoref{sec:conclusion} draws conclusions about our contributions.

\section{System Model} \label{sec:system_model}
In this work, we consider a peer-to-peer network consisting of $N$ peers with open participation. We assume communication channels are established in a decentralized manner, with each peer available for discovery and connection. Furthermore, the channels between peers are unreliable and unordered, similar to the UDP communication model. Consequently, there is no upper bound on message arrival times, and outbound messages may not successfully reach their intended destinations at all. In addition, each peer possesses a cryptographic keypair, of which the public key uniquely identifies the peer within the network and the private key is used to cryptographically sign outgoing messages. Peers interact with each other by exchanging messages according to a reference protocol, such as the Bitcoin network protocol.

\begin{figure}[t]
    \centering
    \begin{subfigure}[b]{0.49\linewidth}
        \centering
        \includegraphics[width=\linewidth]{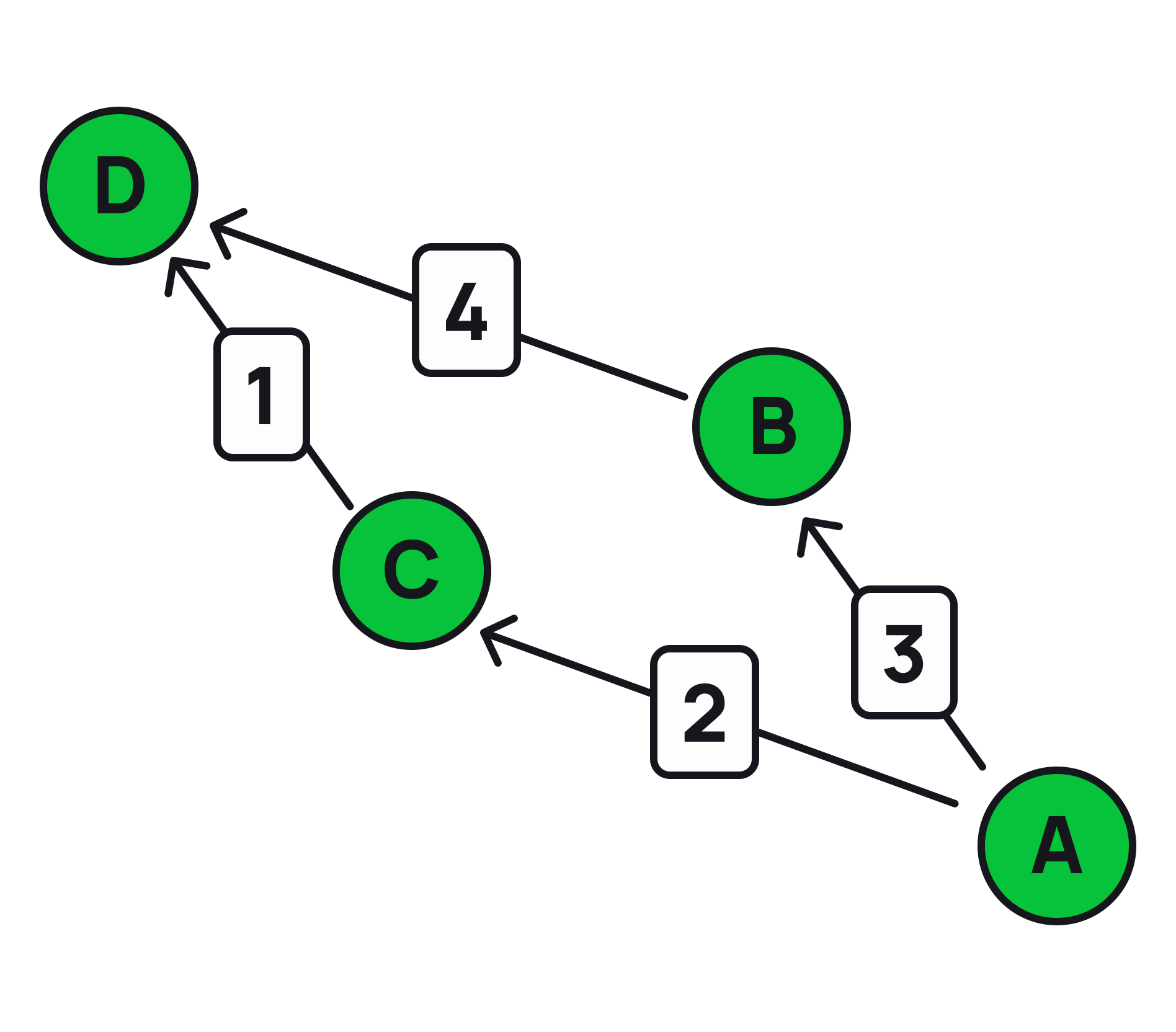}
        \caption[]%
        {{}}    
        \label{fig:example_graph_a}
    \end{subfigure}
    \hfill
    \begin{subfigure}[b]{0.49\linewidth}  
        \centering 
        \includegraphics[width=\linewidth]{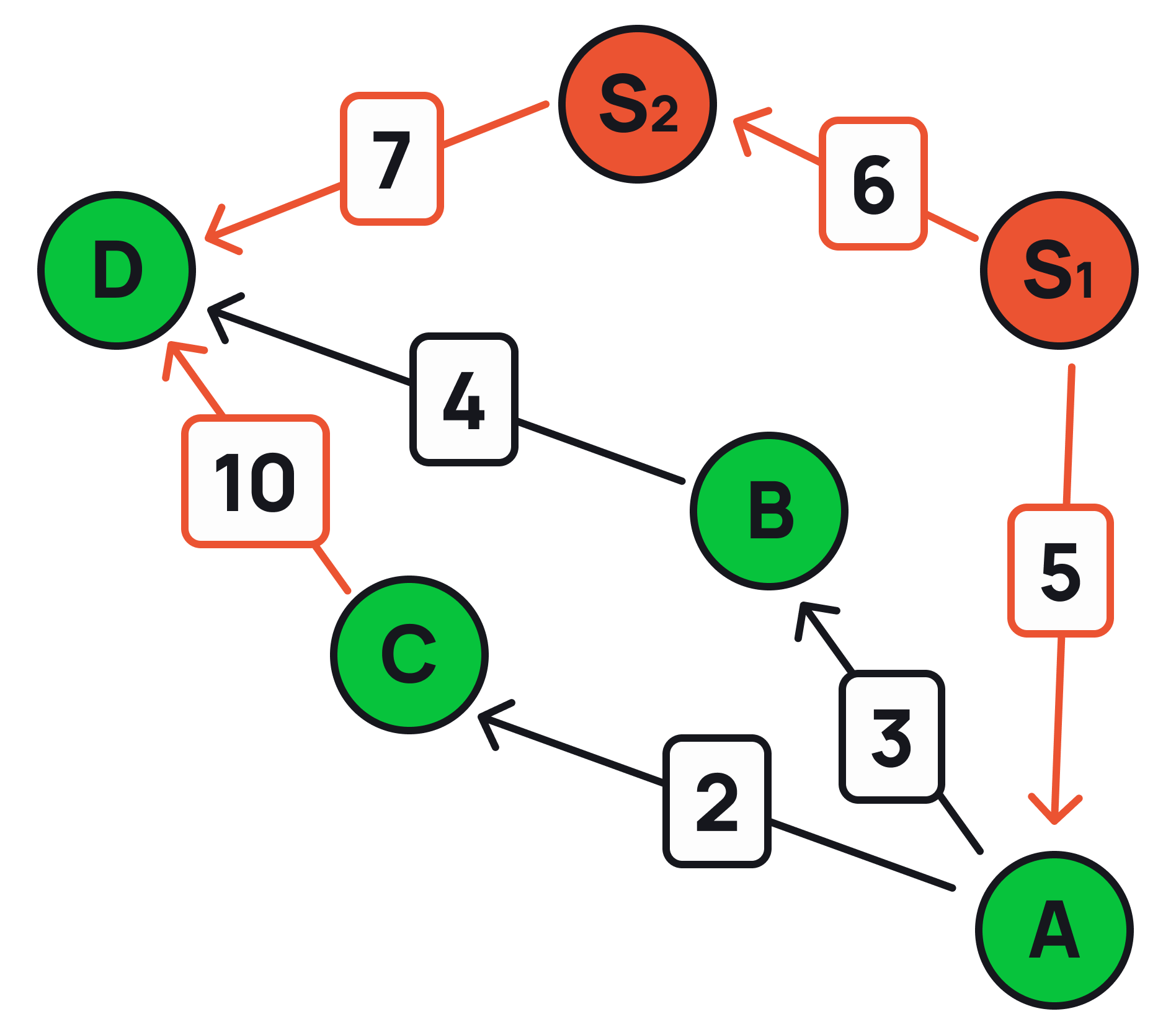}
        \caption[]%
        {{}}    
        \label{fig:example_graph_b}
    \end{subfigure}
    \caption{An example of a contribution graph $G$ (a) and (b) a modified graph $G'$ under a misreport attack with fake contribution edges and Sybil identities $S_1$ and $S_2$ (in red).}
    \label{fig:example_graph}
\end{figure}

\subsection{Problem Description}
We consider the challenge of constructing a universal protocol to maintain network cooperation despite the presence of peers exhibiting selfish behavior, referred to as \textit{selfish peers}. These peers attempt to optimize their personal gain by reaping the benefits of a system while providing little or no actual value to it, thus undermining the cooperative foundation of a peer-to-peer system.

To mitigate selfish behavior, we consider the contributions made to the network by peers, referred to as \textit{utility}. However, due to the inherent complexity of the problem, relying solely on globally verifiable and accurate proofs of each peer's usefulness in a globally-spanning peer-to-peer network is infeasible. Thus, to preserve the decentralized nature, we rely on aggregating \emph{subjective acknowledgments} of utility generated by peers in the network.

The subjective contributions of peers are modeled as a directed weighted graph referred to as a \emph{contribution graph} $G = (V, E, w)$, consisting of a set of nodes $V$ and a set of edges $E$. Each edge $e_{ij} \in E$ is associated with a weight $w_{ij}$ defined by the function $w:V\times{}V\rightarrow{}\mathbb{R}_{\geq{}0}$. A weight $w_{ij}$ corresponds to the cumulative contributions made by $v_j$ that are helpful to $v_i$, where $\{v_i,v_j\} \subseteq V$. An example of a contribution graph can be seen in~\autoref{fig:example_graph_a}, depicting the scoring assigned to nodes based on the subjective contributions of utility.

Using this definition, we can define a cooperation score $C^{G}_k \in \mathbb{R}_{\geq{}0}$ for each $v_k \in V$. This score quantifies the utility of peer $v_k$ with respect to the entire network, created through an aggregation of all cumulative contributions performed by peer $v_k$ for any other peer, i.e., ${w_{lk}~|~l \in V}$. This score is used in mechanisms that rely on global trust scores and can be used to punish selfish peers.

\textbf{Punishing Selfish Peers.} In order to maintain the cooperative nature of the peer-to-peer network, it is essential to identify and punish selfish peers. To achieve this, we introduce a punishment mechanism that relies on the cooperation score calculated for each peer. A selfish peer in the graph $G$ is punished if their cooperation score $C_k^G$ falls below a certain threshold value, denoted by $\tau$. This approach introduces, however, its own set of challenges.

\begin{figure*}[b]
	\centering
    \includegraphics[width=.99\linewidth]{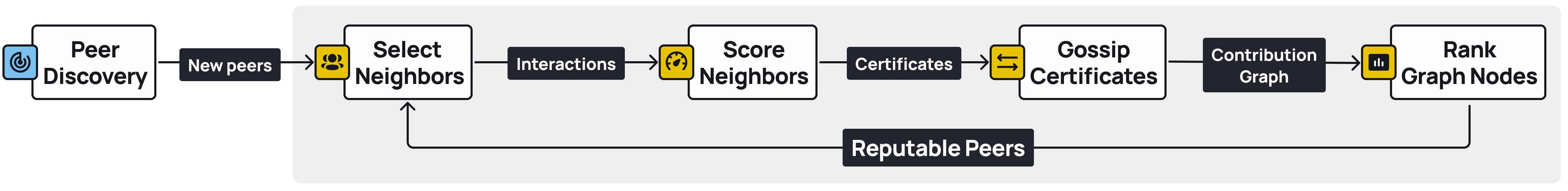}
	\caption{The approach underpinning our mechanism.}
\label{fig:system_model}
\end{figure*}

\textbf{Misreporting and the Sybil Attack.} 
A naive way to aggregate all subjective acknowledgments (e.g., the average of all weights $w_{ij}$ corresponding to a peer $v_j$) is susceptible to misreporting. Misreporting refers to deceptive behavior where participants provide false information about their contributions, seeking an unfair advantage. In conjunction with Sybil attacks~\cite{douceur2002sybil}, strategic peers may attempt to fake their contributions to maximize rewards. Even when employing access restriction techniques, strategic peers can still collude and falsely acknowledge non-existent contributions.

Incorporating the possibility of fake contributions with Sybil peers, we introduce a modified graph $G' = (V', E', w')$, where $V' = V \cup V_s$ denotes a new set of nodes containing a set $V_s$ of Sybil identities, $E' = E \cup E_s$ a set of directed edges with $E' \subseteq V' \times V'$. The modified function $w':V'\times{}V'\rightarrow{}\mathbb{R}_{\geq{}0}$ additionally contains the fictitious utility performed by Sybils and their colluding nodes. Note that, thus, the modified graph $G'$ contains in addition to all nodes and edges in the original graph $G$, Sybils nodes $V_s$ and misreported fake contributions edges $E_s$, created either by nodes together with their Sybils, or with other colluding nodes. A schematic example of this is illustrated in~\autoref{fig:example_graph_b}, depicting the addition of fictitious contributions in a contribution graph. As such, in order to construct a robust protocol that can withstand the challenges posed by misreporting and Sybil attacks, we introduce a requirement for \textit{misreport resistance}.

\textbf{Misreport Resistance.} The goal of this requirement is to identify and punish selfish peers based on their cooperation scores in the modified graph $G'$, even in the presence of fake contributions and Sybil identities. Let $C_k^{G'}$ be the cooperation score of node $v_k$ in the modified graph $G'$. A selfish peer is punished in the graph $G'$ if their cooperation score $C_k^{G'}$ falls below a certain value $\tau'$, where the difference between $\tau'$ and the original threshold $\tau$ is bounded by a predefined error margin $\epsilon$, i.e., $|\tau - \tau'| \leq \epsilon$.

By incorporating this misreport resistance requirement, we aim to devise a protocol that can effectively maintain the cooperative nature of the peer-to-peer network, even in the face of adversarial behavior involving misreporting and Sybil attacks. The proposed approach seeks to accurately evaluate the contributions of peers while minimizing the impact of fake contributions and ensuring that strategic peers cannot gain an unfair advantage over their honest counterparts.

\subsection{Selfish Peer Starvation Approach}

A key aspect of our approach to handling selfish peers is the reliance on the subjective estimation of cooperation scores. Each peer independently evaluates the cooperation score of other peers based on their direct and indirect contributions. We refer to the subjective cooperation score of peer $v_i$ from the point of view of the evaluating peer $v_j$ as $C_{ij}^G$. 

This subjectivity effectively limits the impact of misreporting since a peer is considered cooperative only if it is directly or indirectly helpful to the evaluating peer. Direct utility is represented by the direct edges originating from the evaluating peer in the contribution graph. Indirect utility, on the other hand, is manifested through the existence of a path between the evaluating peer and the peer being evaluated in the contribution graph. Selfish peers will typically have fewer connections,
and the few paths leading to them will have lower weights. Their ability to misreport is also severely limited, as they must perform honest work to receive acknowledgments and create paths in the contribution graph.

Our approach, however, also presents a potential downside: it may overlook the genuine work performed by two honest peers that have no connection to the evaluating peer in the cooperation graph. To mitigate this issue and improve network connectivity, honest peers actively participate in numerous interactions throughout the network, thereby ensuring that the honest peers in the cooperation graph remain connected.

As the network evolves, peers gradually begin to favor more reciprocal partners for future interactions. This shift in behavior leads to a natural consequence for selfish peers, who end up with few or no edges, effectively limiting the services they can access. Sybil attacks also become impractical, as the malicious nodes end up in an isolated subgraph, disconnected from other honest peers.

\section{Solution Overview} \label{sec:solution}
We present a practical protocol designed to sustain cooperation in any peer-to-peer network. The primary objective of this design is to achieve the eventual isolation of selfish peers while ensuring that the network remains connected, allowing honest peers to communicate.

\subsection{Overview}

Each peer in our protocol evaluates the relative contributions of its overlay neighbors. Peers are ranked according to their relative contributions to the network. For instance, peers might be ranked based on the utility they provide for a gossip protocol or how reliably they share files with other peers. Hence, the types of utility taken into account are protocol-dependent.

Our design is composed of four mechanisms, which can be summarized to \textit{Select Neighbors}, \textit{Score Neighbors}, \textit{Gossip Certificates}, and \textit{Rank Graph Nodes}.~\autoref{fig:system_model} visualizes these processes and their interactions within our system. These four mechanisms work together to adjust the peer-to-peer overlay, accounting for selfish peers. Our mechanism proceeds in rounds. As we operate in a fully decentralized setting, the scheduling and the length of each round are up to individual peers. This round-based approach ensures liveness by connecting with still-unknown peers and lower-rated peers. The details surrounding the selection of these connections are discussed later on in this section.
As the first process, after peer discovery, relies on prior knowledge generated through past interactions, as depicted by the feedback loop visible in~\autoref{fig:system_model}, we start by outlining the \textit{Score Neighbors} mechanism.

\textbf{Score Neighbors}
Each peer interacts with its locally connected peers, for instance by receiving or sending messages according to some reference protocol. After these interactions, peers store the information about the provided utility. For example, in a file-sharing scenario, the utility might be the number of files shared or the amount of data transferred, with an associated weight reflecting the importance of that contribution. Peers locally store the cumulative utility provided in the form of a \textit{signed certificate}, showing that some peer $v_j$ was helpful to peer $v_i$ with a utility of $w_{ij}$. This certificate is signed by $v_i$ and shared with $v_j$.

\textbf{Gossip Certificates}
At the same time, each peer periodically runs a pull-based crawler to collect signed certificates from their neighbors. Thus, neighboring peers exchange information about each other by sharing the latest known certificates. Note that merely storing the most recent certificates is sufficient, as they reflect the latest cumulative opinion on the subjective utility provided by a peer.

\textbf{Rank Graph Nodes}
Based on the collected and created certificates, each peer can reconstruct part of the contribution graph. For each node in the reconstructed graph, we assign a cooperation score. This is done with the help of the personalized reputation function MeritRank~\cite{nasrulin2022meritrank}.

\textbf{Select Neighbors}
The peers are ranked according to their reputation score. Further interactions within the network are adjusted based on these rankings. Peers prioritize interactions with high-ranking peers while filtering out peers with consistently low reputations. As a result, peers can sustain cooperation in the network, effectively isolating selfish peers as a punishment and promoting collaboration.

\subsection{Create and Gossip Certificates}
As peers interact with each other, they evaluate the provided utility. It is important to note that our approach is agnostic to the specific method by which a peer evaluates another peer. However, we do assume that at each round, peers record their evaluation in the form of a numerical value. As mentioned, the utility $w_{ij}$ provided by peer $v_j$ to peer $v_i$ is recorded as a contribution certificate $\mathtt{cert_{ij}}$, which can be defined as:

\begin{align*}
\mathtt{cert_{ij}} = (pk_i, pk_j, w_{ij}, r_i, sign_i)
\end{align*}

A tuple denoted by $\mathtt{cert_{ij}}$ represents a record of utility $w_{ij}$ provided by peer $v_j$ for the benefit of peer $v_i$. It includes parameters such as $pk_i$ and $pk_j$, which serve as identifiers for peers $v_i$ and $v_j$, respectively. Additionally, the round number $r_i$ is used as a unique counter to show the local round in which the work was completed. Finally, the $sign_i$ parameter represents the cryptographic signature generated using the private key of peer $v_i$ to ensure the authenticity and integrity of the record.

Periodically, peers pull random certificates from their overlay neighbors. Once a certificate for $w_{ij}$ is received, a peer $v_k$ checks its last known certificate for $(i, j)$ and replaces it with the new one, updating the edge weight $(i, j)$ of its locally known contribution graph. Over time, peers will accumulate more certificates, enabling them to construct a more accurate representation of the contribution graph. 

\subsection{Sybil-Tolerant Reputation Ranking}
In the local contribution graph, each identifiable node is assigned a ranking that corresponds to its relative contribution to the network. Straightforward approaches that rely on basic global centralized measures, such as total network contribution, are highly susceptible to manipulation. Furthermore, even sophisticated global reputation mechanisms, such as PageRank~\cite{pagerank1998}, are in principle not Sybil-tolerant~\cite{cheng2005sybilproof}. Consequently, malicious actors can exploit this vulnerability to accrue an undeserved reputation by misreporting the contributions made in conjunction with Sybils.

\begin{figure}
    \centering \includegraphics[width=.9\linewidth]{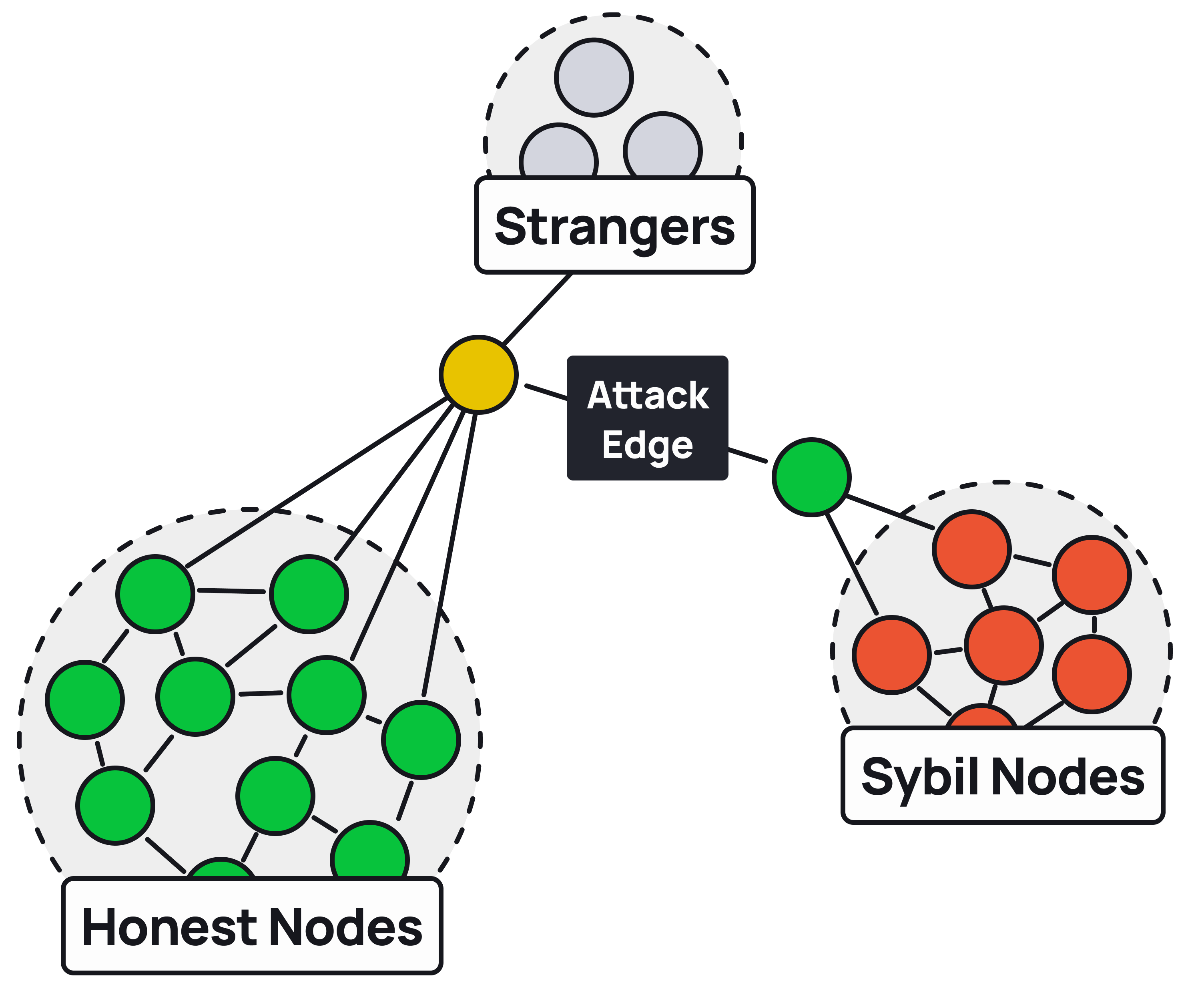}
    \caption{An example of a connection graph.} \label{fig:selection_graph}
\end{figure}

There exist two types of Sybil attacks: an \textit{active} and a \textit{passive} variant~\cite{sybil_proofness_2021}. In the active Sybil attack, any Sybil may have a direct edge to honest peers. Whereas in the passive Sybil attack, there exists a single node, connected to honest peers, with which the Sybils have a direct edge. This latter variant can be seen in~\autoref{fig:selection_graph}, portraying the connections from the point of view of the node highlighted in yellow. The edge labeled \textit{attack edge} is an instance of a passive Sybil attack. The colluding node in green has numerous connections to the cluster of Sybils with fictitious trust scores. 

Personalized reputation mechanisms tackle this problem by attributing a positive reputation to node $v_j$ solely if a path exists from a seed node $v_i$. This feature intrinsically offers resistance to Sybil attacks, as participants are required to execute tasks to establish a path. We base our reputation algorithm on MeritRank~\cite{nasrulin2022meritrank}, which is a Sybil-tolerant personalized reputation mechanism. The algorithm is based on random walks starting from a source node. The personalized cooperation score of a node $v_j$ from the perspective of source node $v_i$ on a graph $G$ is the steady-state probability of an $\alpha$-terminating random walk reaching a node $v_j$. The parameter $\alpha$ is a transitivity decay, that limits the length of random walks. In practice, we initiate $R$ random walks from an individual node within the cooperation graph. The cooperation score of the $C^{G}_{ij}$ algorithm is determined as the fraction of random walks that start from node $v_i$ and pass through node $v_j$. 


The Sybil tolerance of MeritRank arises from two design choices: the relative personalized ranking and the transitivity decay. The scores are determined based on relative contributions and are periodically recalculated. The score of nodes that stop contributing will naturally decay in comparison to nodes that consistently contribute. This occurs due to the inflation of the global utility generated by nodes within the system. A Sybil attack becomes unfeasible as it necessitates constant genuine contributions in order to obtain positive scores from honest nodes. Transitivity decay, on the other hand, limits the influence of nodes that are distant from the seed node. By incorporating the parameter $\alpha$, random walks are more likely to terminate as they traverse further from the seed node. Consequently, the reputation score reflects the genuine contribution of a node to the network, as observed from the seed node's standpoint.

This approach prevents Sybil nodes from accumulating high reputation scores by merely connecting to other Sybil nodes or manipulating the structure of the network. 

\subsection{Peer Selection} 
\begin{figure}
	\centering \includegraphics[width=.8\linewidth]{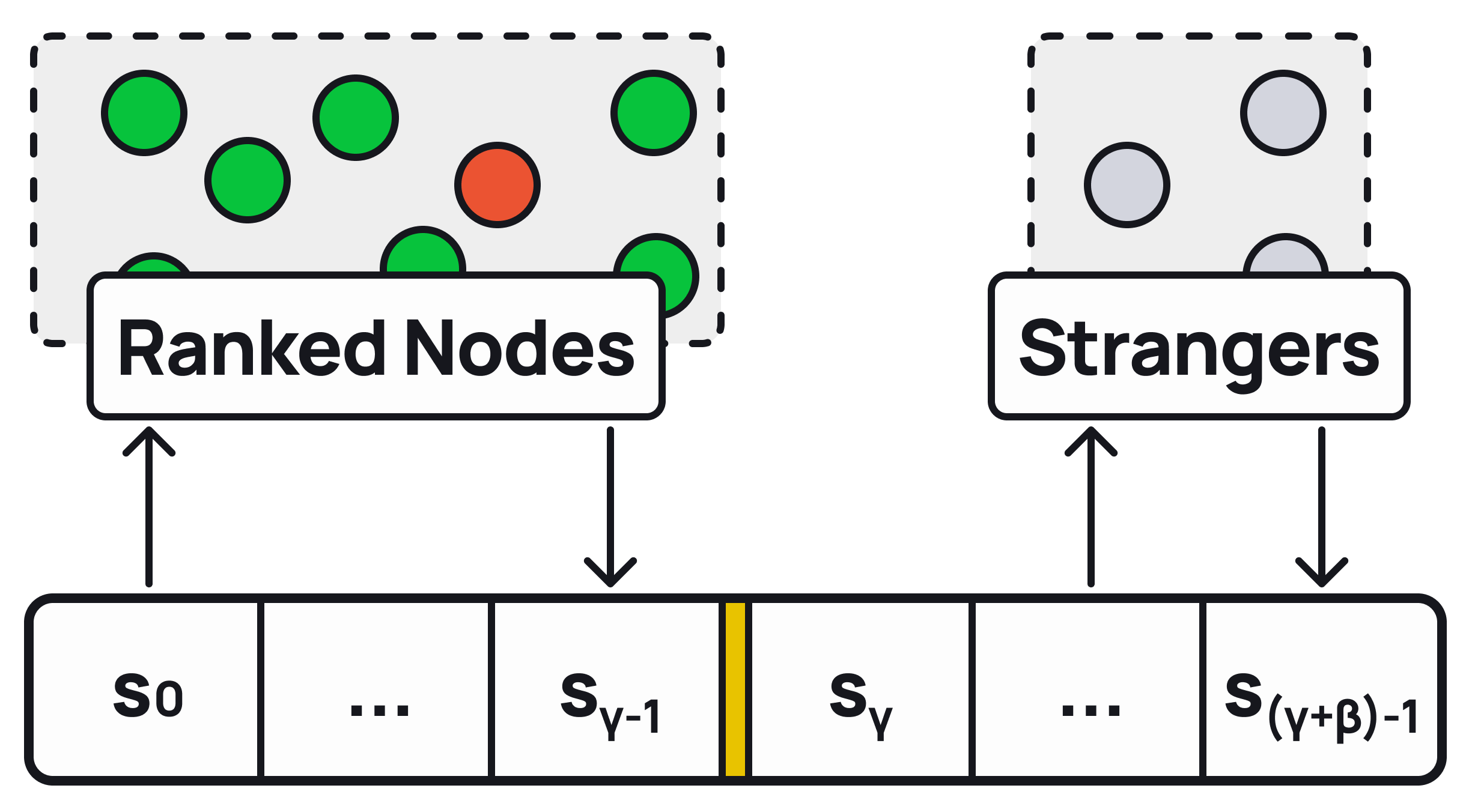}
	\caption{The peer selection mechanism with $\gamma$ slots for reputable peers and $\beta$ slots for stranger peers.}
\label{fig:selection_slots}
\end{figure}

The peer selection process consists of several stages, with each peer maintaining a dynamic list of selected peers for primary communication during the current round. We employ a specific notation for peer selection, utilizing $n$ slots: a fraction of $\gamma$ slots are reserved for reputable peers and a fraction of $\beta$ slots for stranger peers with zero reputation, where $\gamma + \beta=n$. The slots are filled using push and pull strategies, as illustrated in~\autoref{fig:selection_slots}, and are shuffled after each round $r_i$.

Initially, a peer proactively fills a predetermined fraction of slots based on the set of known reputable peers and stranger peers, a process referred to as the pull strategy. The peer sends a request to the selected peer, occupying the push slot of the recipient peer. The remaining slots are reserved for the push strategy and are filled by incoming requests. If all push slots are occupied, the incoming request will be rejected. This leads to a system in which both highly-ranked and lower-ranked nodes have the ability to form connections. This counteracts a situation in which only connections to highly ranked nodes are made, as this would lead to an inability for new or low-ranked nodes to gain or improve their reputation.

\textbf{Bootstrap Period}.
When a new client joins the network, this node maintains a dynamic list of potential peers, which may be identified through various methods such as querying a centralized directory, distributed hash tables, or gossip protocols. During the bootstrap period, peers thus do not account for reputation to ensure the network remains connected. Specifically, during this period, $\gamma = 0$ and $\beta = n$. Upon the conclusion of the bootstrap period, $\gamma$ increases and $\beta$ decreases correspondingly, to account for contributions of nodes documented in the meanwhile.

A typical connection graph is visualized in the aforementioned~\autoref{fig:selection_graph}. This visualization is performed from the point of view of the node highlighted in yellow. In a typical scenario, a node will favor connections to honest, thus highly reputable nodes. This is depicted by the high number of connections to that respective cluster. However, strangers must also be given the opportunity to create a reputation for themselves in order to guarantee liveness. As such, our node also has connections with strangers. Finally, this figure also depicts how Sybils can still occur in the system, though they are of minor impact as they require a peer that is reputable to the peer central in this figure, hence, there must still be utility created. Furthermore, their score is bounded by said utility. As such, this visualizes how our system is Sybil-tolerant, as Sybils will never be trusted more than the utility provided by their strongest link through a colluding node. This is visualized by the \textit{attack edge}.





\section{Experimental results} \label{sec:results}
We provide two experimental setups to evaluate our proposals. First, we simulate a blockchain network based on the latest observed data from the Bitcoin peer-to-peer network. Second, we deploy our accounting mechanism to a client of the file-sharing network of the BitTorrent client Tribler~\cite{tribler}. We report on the data crawled from the network.

\subsection{Experimental Setup}

\textbf{Bitcoin use case}.
To demonstrate its practicality, we have applied the mechanism defined in~\autoref{sec:solution} to a simulation of the Bitcoin network. We have simulated numerous aspects of the Bitcoin network in order to benchmark our solution in a realistic setting. The code for this simulation in Python can be found in our repository\footnote{\url{https://github.com/tribler/bami}.}. We define the utility of a Bitcoin miner in its ability to deliver non-spam transactions and non-stale, non-orphan blocks. Non-spam transactions are transactions that are valid and have a non-zero fee amount. A node $v_i$ creates a certificate for a peer $v_j$ if $v_i$ receives a useful transaction or block from $v_j$.   

We model the experiment after a topology as discussed in~\cite{grundmann2022peer}, which shows that on average 7518 Bitcoin nodes are reachable, and after peer-to-peer latencies as reported in~\cite{rohrer2021blockchain}. As such, we use $N=7518$ for our Bitcoin experiments unless specified otherwise. For MeritRank, we fix the parameter of transitivity decay to $\alpha=0.2$ and $2000$ random walks. A Bitcoin transaction is created at some miner node and then relayed over the network to other miner nodes. At the same time, miners run a block creation with an average of 10 minutes by selecting transactions from a local mempool. We simulate a Bitcoin network for 30 days with a transaction rate of 2.5 transactions per second, with all nodes progressing to the next round $r_i$ after each day.

\textbf{Tribler use case}.
As free-riding is a well-researched issue in BitTorrent, we have applied an implementation of our accountability mechanism to the Tribler~\cite{tribler} network. Tribler encompasses all logic of the BitTorrent protocol in the form of a file-sharing network.
The utility of a peer in Tribler is defined as the amount of bandwidth delivered for the file-sharing and bandwidth-sharing service.

\subsection{Selfish Peers}


\begin{figure}
\centering
\includegraphics[width=0.9\linewidth]{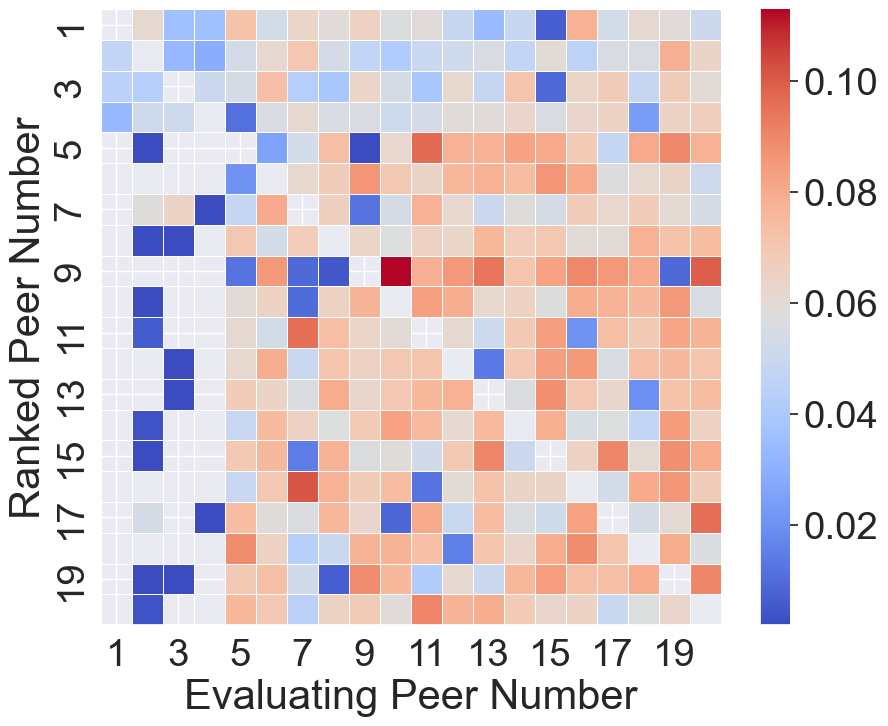}
\caption{A heat map with ranking scores (N=20). Nodes 1-4 are selfish with a share ratio of 0.2.}
\label{fig:heatmap}
\end{figure}

We simulate the Bitcoin network with injected selfish peers. We test different \emph{share ratios} for the selfish peers: a factor between 0 and 1 representing the amount of data they share. An honest peer has a share ratio of 1.

\begin{figure}
\centering
\includegraphics[width=0.9\linewidth]{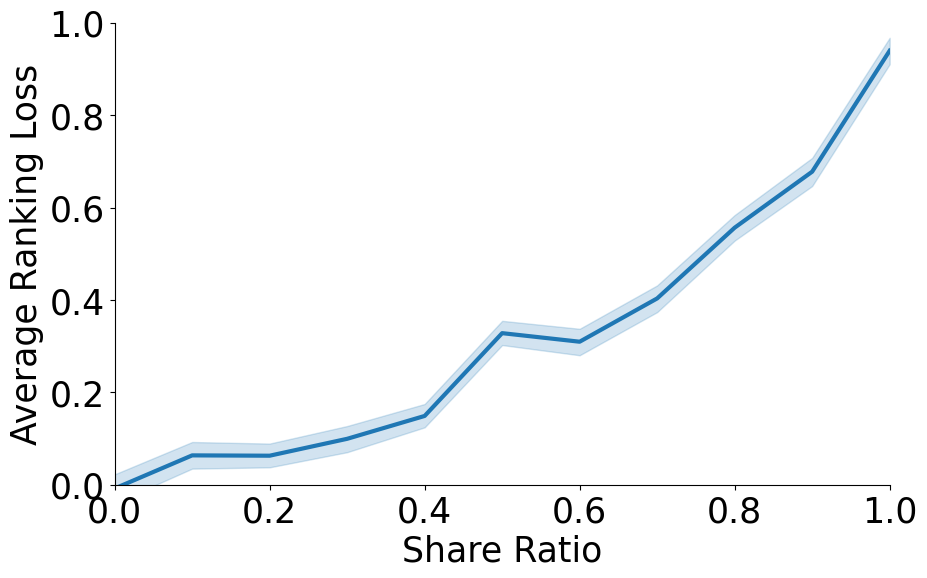}
\caption{Average ranking loss of selfish nodes (N = 7518, with 1800 selfish peers).}
\label{fig:rank_ratio}
\end{figure}

\autoref{fig:heatmap} reports a heat map matrix of the subjective ranking of each peer. We use a small network (N=20) for demonstration purposes. The peers with ids 1-4 are selfish sharing only 20\% of the data known to them. The selfish nodes are less preferred than any other nodes by the honest peers, with many selfish peers identified simply as strangers.

\begin{figure}
    \centering        
\includegraphics[width=0.9\linewidth]{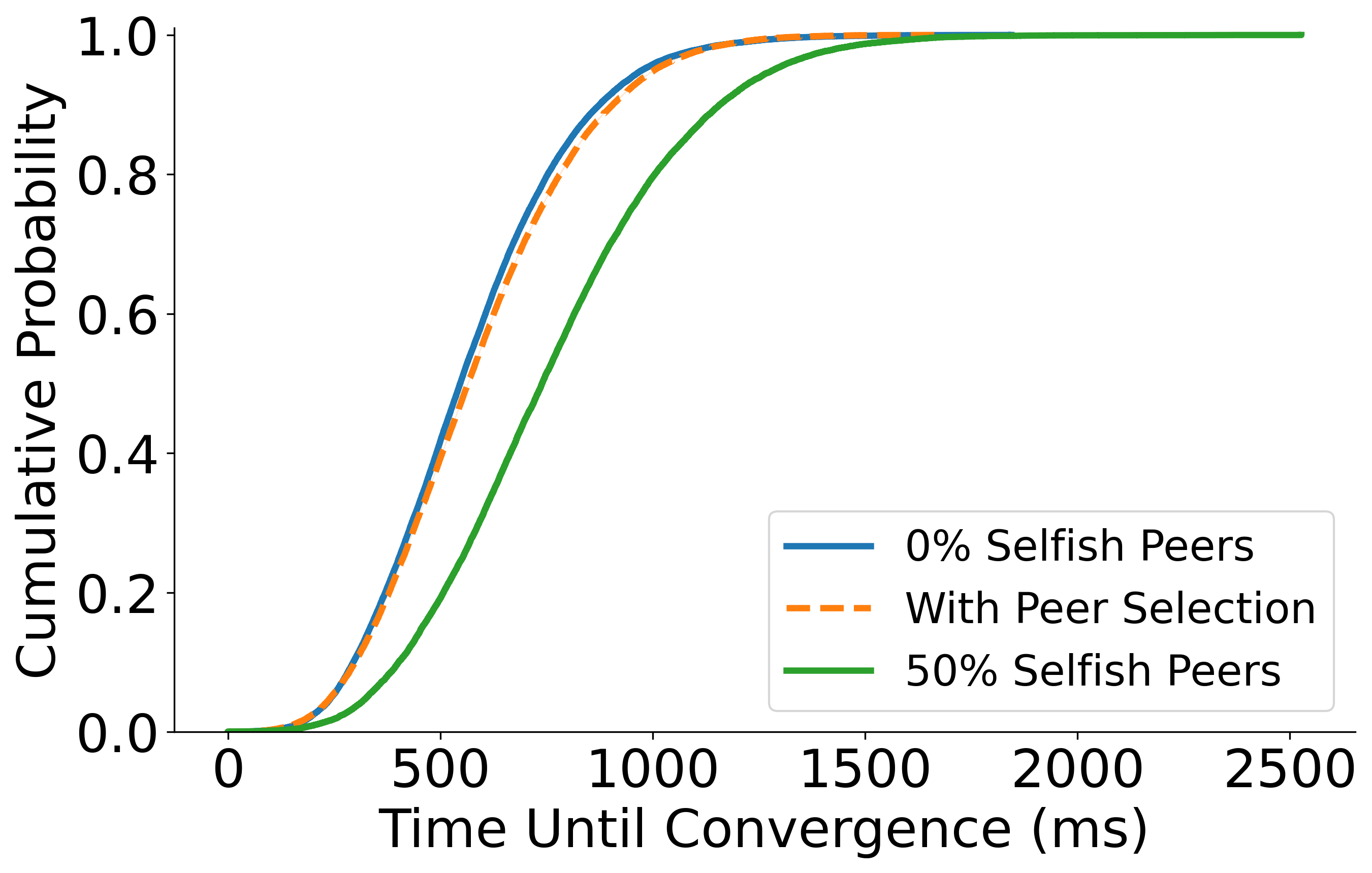}
    \caption{The effect of selfish nodes on convergence within Bitcoin.}
\label{fig:convergence}
\end{figure}

We vary the \textit{share ratio} parameter for the selfish peers. We run the simulation with 25\% of the network (i.e., 1800 peers) being a selfish peer, each sharing only part of the mempool and blocks. We use the average ranking loss, i.e., the ratio of the average ranking of selfish peers to the average ranking of non-selfish peers. This metric is used to signal the degree to which we can successfully detect and punish selfish peers. We report our findings in~\autoref{fig:rank_ratio}: selfish peers suffer an immediate loss in ranking when they share 10\% less than the average honest peer.

\autoref{fig:convergence} demonstrates the effect of selfish peers on data convergence. Unsurprisingly, selfish peers can delay the network convergence, delaying each transaction on average by 400 ms. When a peer selection mechanism is employed with $\gamma = 100$ with $n=125$, we manage to negate the effect of selfish peers with an average delay of 20 ms.

\subsection{Tolerance to Sybil Attack}

\begin{figure}
\centering  \includegraphics[width=0.9\linewidth]{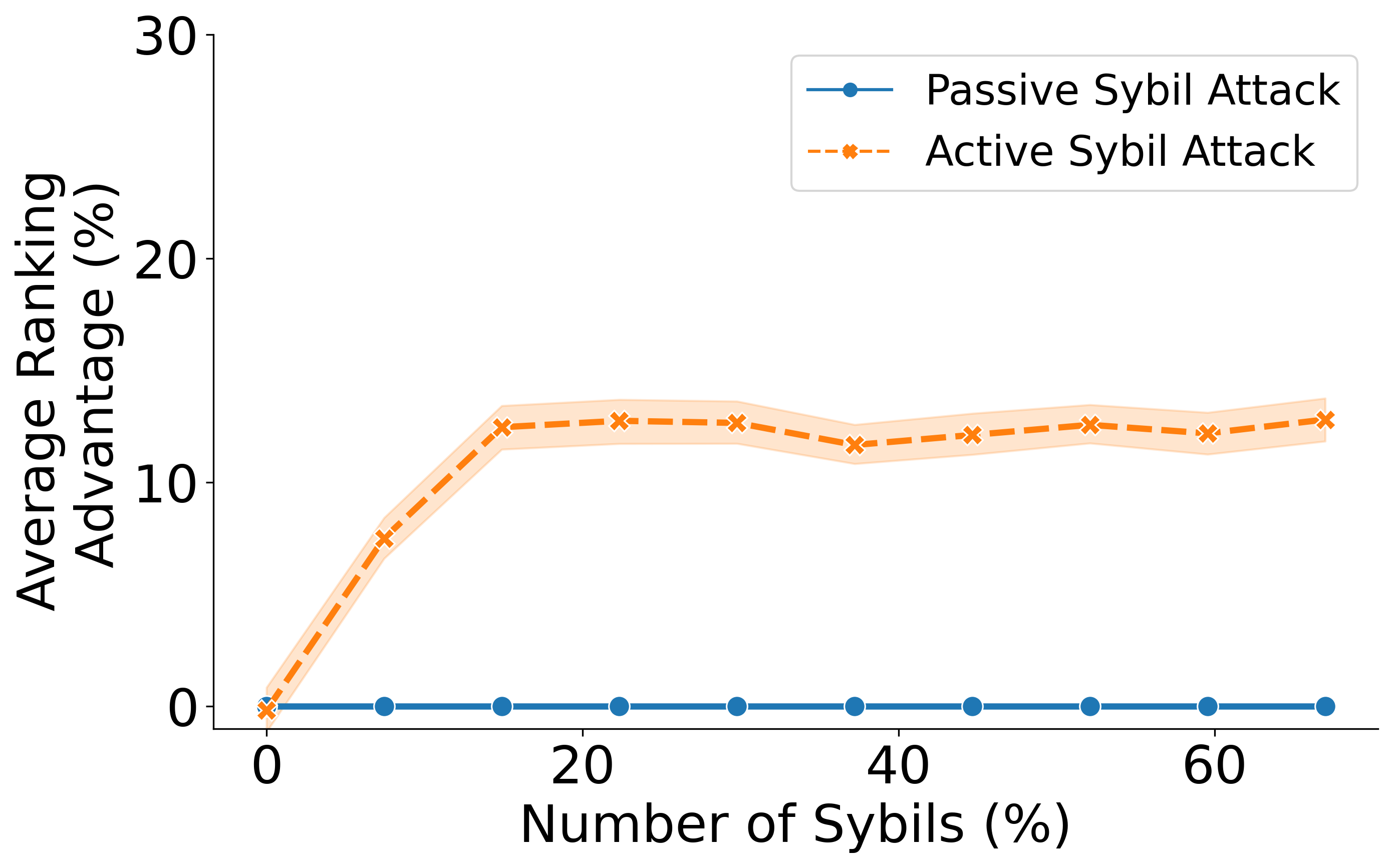}\caption{The effect of Sybil attacks on the ranking.}\label{fig:sybil}\end{figure}

To demonstrate the effect of the Sybil attack, we create a Sybil region attempting to manipulate the ranking of one attacker peer. We implement two attack strategies, a passive Sybil attack where each new Sybil peer creates a certificate with one peer and an active Sybil attack where each Sybil creates certificates with each other and the attacker peer. The passive attack can be executed by spamming the peer discovery and introducing fake peers; however, an active attack requires active maintenance of all Sybils to be able to connect to other honest peers.

\autoref{fig:sybil} shows the effect of Sybils on the ranking of selfish peers in the network. The x-axis represents the percentage of Sybils in the network, ranging from 0\% to 67\%. The y-axis shows the average ranking gained, expressed as a percentage. For the active Sybil attack, we report the total advantage gained by all Sybil peers. The passive attack has no effect on the ranking of selfish peers, while the active variant increases the ranking advantage of selfish peers up to a point. Specifically, the selfish peer together with its Sybils is preferred at an average 12\% higher than deserved. 

\begin{figure}
\centering  \includegraphics[width=0.9\linewidth]{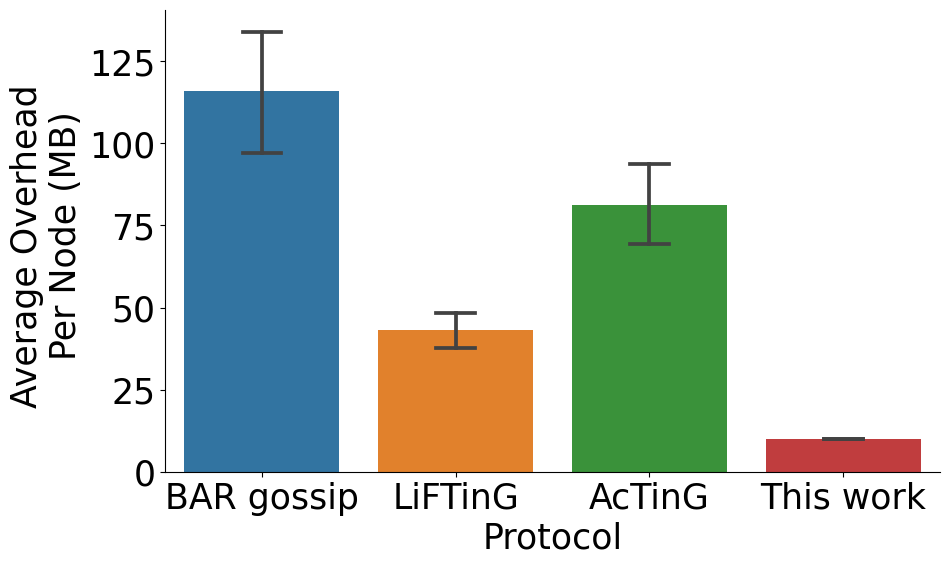}\caption{The overhead introduced by the protocols.}\label{fig:overhead}\end{figure}

The result shows that nodes will not gain any advantage by naively manipulating the ranking through misreporting. To gain any advantage, nodes need to execute a mass-scale attack, effectively maintaining a subnetwork of Bitcoin peers.

\subsection{Overhead}
We introduce only marginal overhead on the bandwidth and memory. One certificate is 220 bytes; as such, after a month's time, a network of $N=7518$ requires the storage and exchange of up to 10 MBytes of certificates. This is enough to ensure reasonable connectivity and a correct run of the Bitcoin peer.

The number of walks used for MeritRank controls the accuracy of the ranking estimation in exchange for performance. However, the choice of 2000 walks gives an accurate enough estimate and requires only 100 milliseconds for a graph with $7518$ nodes.   

\begin{figure}[b]
    \centering
\includegraphics[width=0.85\linewidth]{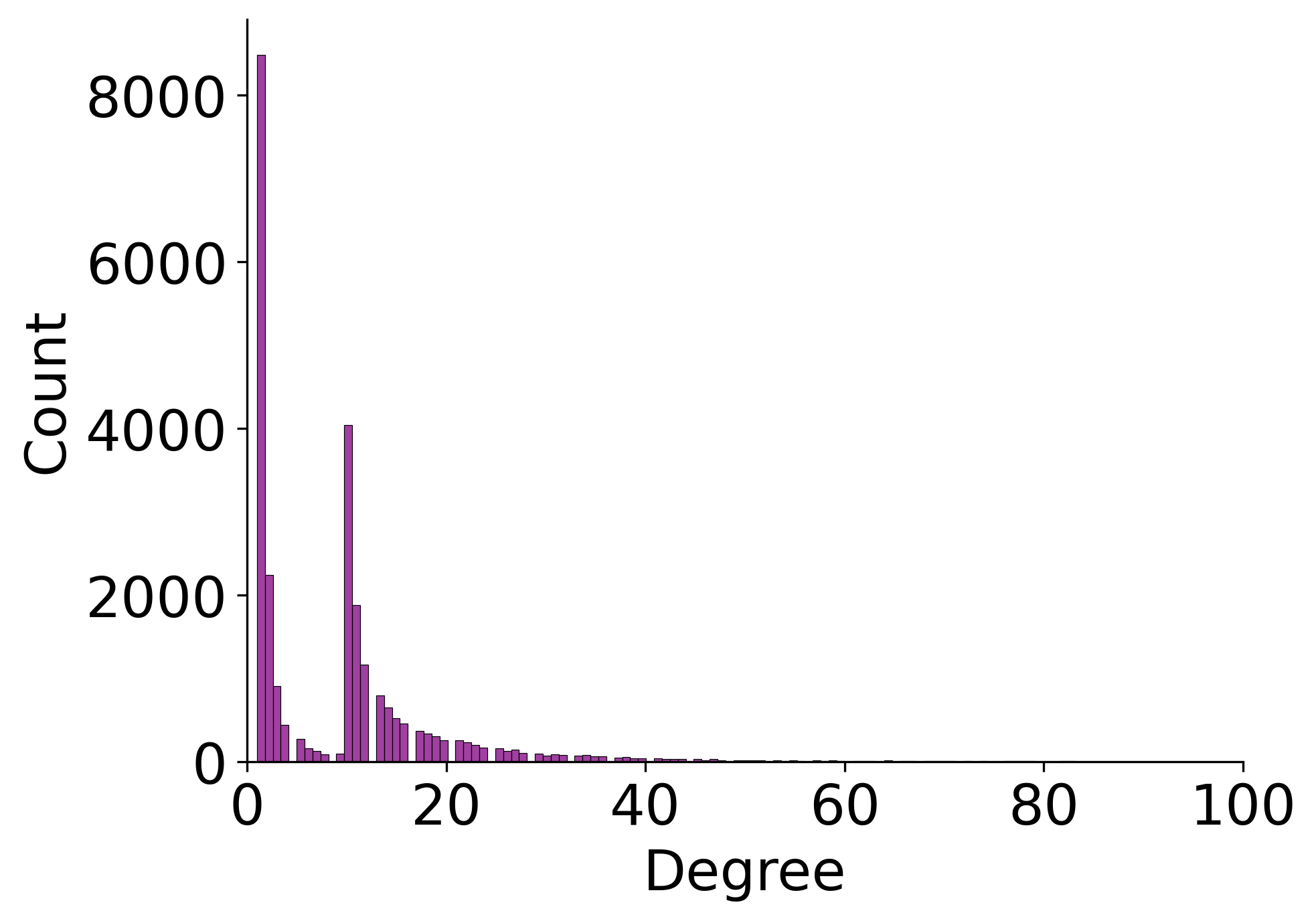}
    \caption{Degree distribution of Tribler contribution graph.}
    \label{fig:tribler_degree}
\end{figure}

We conducted a comparative analysis of our protocol against related work, namely BAR gossip\cite{li2006bar}, LiFTinG\cite{lifting2010}, and AcTinG\cite{acting2014}. In our simulations, the selfish nodes had a share ratio of 0.2. We set the fanout to 8 with a gossip period of 500 ms. For LiFTinG, we fixed the number of monitors for scorekeeping at $M=25$. In the case of AcTinG, partners audited each other during the connection with a probability of 15\%, in accordance with the recommendation in the original paper~\cite{acting2014}.

We simulated the protocols using the Bitcoin dataset and measured the average bandwidth overhead per node over 10 runs, as shown in~\autoref{fig:overhead}. Both BAR gossip and AcTinG resulted in significant bandwidth overhead due to the necessity of sharing and auditing the entire history log. LiFTinG, on the other hand, conserved bandwidth by only cross-validating the message. Our protocol introduced the least overhead, as it only required periodic certificate exchanges to detect selfish nodes.

\subsection{Tribler Deployment Results}
~\autoref{fig:tribler_degree} presents the degree distribution of the contribution graph, which was collected over a year-long period from April 2022 to April 2023. The contribution graph comprises 27,259 nodes and 204,927 edges. The majority of peers interacted with no more than 10 peers, with a median of 1 peer. However, we identified 781 peers with a degree exceeding 100. A small number of peers exhibited an exceptionally high degree of 446 and shared 13 TB of data. These observations align with previous reports of a reliance on a small number of altruistic nodes~\cite{piatek2007incentives}.

We observed that many peers have only a few connections to each other, without forming a connected component. Specifically, the largest connected component comprises 7382 nodes. However, due to the natural skewness of the contribution and the properties of MeritRank, a node can rank only between 600 and 1000 nodes on average. This observation highlights a potential limitation of our approach, suggesting the need for interaction with more nodes to enter the ranked connected subgraph. We defer the discussion on incentivizing users to form a bigger connected subgraph to future work.

\section{Related Work} \label{sec:related_work}




\begin{table}[b]
\caption{Comparison with Related Work} \label{tab:related_work}
\resizebox{0.99\linewidth}{!}{
\begin{tabularx}{1.08\linewidth}{lcccc}
\toprule
        & \multicolumn{1}{c}{\textbf{\begin{tabular}[c]{@{}c@{}}Sybil\\ Tolerant\end{tabular}}} & \multicolumn{1}{c}{\textbf{\begin{tabular}[c]{@{}c@{}}Score-Based\\ Reputation\end{tabular}}} & \multicolumn{1}{c}{\textbf{\begin{tabular}[c]{@{}c@{}}Non-\\ Global view\end{tabular}}} & \multicolumn{1}{c}{\textbf{\begin{tabular}[c]{@{}c@{}}No Trusted\\ Third Party\end{tabular}}} \\
\midrule
\textbf{This work}    & \cmark                                           & \cmark                                                   & \cmark                                            & \cmark                                                   \\
LiFTinG~\cite{lifting2010}      & \xmark                                           & \cmark                                                   & \xmark                                            & \cmark                                                   \\
AcTinG~\cite{acting2014}       & \xmark                                           & \xmark                                                   & \xmark                                            & \cmark                                                   \\
PAG~\cite{pag2016}          & \xmark                                           & \xmark                                                   & \xmark                                            & \cmark                                                   \\
PeerReview~\cite{peerreview2007}   & \xmark                                           & \xmark                                                    & \xmark                                            & \cmark                                                   \\
FullReview~\cite{fullreview}   & \xmark                                           & \xmark                                                    & \xmark                                            & \cmark                                                   \\
ConTrib~\cite{devos2021}      & \xmark                                           &  \cmark                                                    & \cmark                                            & \cmark                                                   \\
BAR~\cite{li2006bar}          & \xmark                                           &  \xmark                                                    & \cmark                                            & \xmark             \\                                                                 
\bottomrule 
\end{tabularx}
}
\end{table}

An adjacent research field to this work is that of accountability in decentralized systems. ConTrib~\cite{devos2021} is a mechanism for tracking resource usage across different systems. It prevents abuse by recording resource contributions and consumption in personal ledgers. Parties document these interactions in a cooperative manner: a client proposes a record encapsulating the contribution or consumption of a certain resource in a specific system. The receiver then acknowledges this proposal through a confirmation message. In contrast to our proposed system, ConTrib actively attempts to document abuse and relies on multicasting during both phases of documenting interactions and through pull-based exchanges. As a result, our design reduces the overall message complexity by documenting only merit and does not require agreement.

BAR Gossip~\cite{li2006bar} is a byzantine, altruistic, and rational tolerant gossip protocol designed for live event streaming. Apart from receiving data directly from a broadcaster, clients receive data through \textit{optimistic push} and \textit{balanced exchange}. In the former mechanism, clients push data in the hope that receiving clients will eventually return the favor. In the latter variant, clients both exchange an equal amount of updates. Clients that divert from the protocol are punished through a proof of misbehavior, leading to a subsequent eviction from the network. BAR relies on the sharing of both the private and the public key with the sole broadcaster.

PeerReview~\cite{peerreview2007} proposes a generic accountability mechanism for distributed systems, in which all messages sent and received by a node are recorded to detect deviations from the protocol. Misbehaving nodes can become suspected and subsequently exposed. A node becomes suspected when it does not acknowledge a certain message. It can exonerate itself from suspicion by acknowledging the initial message. Nodes that are exposed for their misbehavior permanently no longer receive new data. Similarly, FullReview~\cite{fullreview} additionally targets selfish nodes through the use of game theory.~\cite{avms2010} proposes a methodology for auditing through Accountable Virtual Machines, relying on the analysis of execution logs. These mechanisms, while highly accurate, impose significant overhead when applied to any type of peer-to-peer system.

LiFTinG~\cite{lifting2010} introduces a protocol designed to detect free-riders in gossip-based systems. This detection is accomplished through two verification mechanisms: direct and a posteriori. In the direct approach, a gossiping node verifies that all transmitted messages are proposed to the correct number of nodes. In the a posteriori approach, clients are asked to submit their log of past interactions, enabling the validation of expected behavior. Nodes that misbehave can be blamed and eventually expelled based on their reputation score, which is managed by assigned managers for each node. AcTinG~\cite{acting2014} proposes a similar system, ensuring zero false positives and addressing the issue of colluding selfish nodes. In AcTinG, all clients record their updates in a secure local log. Other clients may then request these logs, revealing any selfish behavior if data was not forwarded according to the protocol. PAG~\cite{pag2016} suggests a privacy-preserving accountability mechanism. Each node has a set of monitors that verify whether the specific node correctly gossips data to the expected nodes. This verification is carried out through the validators of the receiving nodes. LiFTinG, AcTinG, and PAG specifically target gossip-based content dissemination protocols, making them not directly applicable to all distributed systems.

\autoref{tab:related_work} contrasts our work with the related work discussed earlier, based on four criteria we consider essential for decentralized networks. Sybil tolerance is crucial to counteract attacks where peers manipulate their scores through Sybils. We argue that tolerating selfish nodes is more advantageous than entirely eliminating them. A score-based reputation system, as opposed to a binary one, allows for a more nuanced evaluation and enables diverse punishment strategies. Dependence on a global view (e.g., full membership) is impractical for permissionless systems and, depending on the size, may be unfeasible. Therefore, we favor solutions that operate without requiring full network visibility or full network reachability. Lastly, reliance on trusted third parties introduces potential attack vectors, leading to privacy concerns and collusion risks. As shown, our work is the first to satisfy all of these criteria.

\section{Conclusion} \label{sec:conclusion}
We have presented our protocol for sustaining cooperation in peer-to-peer networks. Our work can function in any type of peer-to-peer system and isolates selfish peers who abuse the network. It does so while introducing minimal overhead and minimizing the risk posed by Sybils. 
The crucial aspect of our design is \emph{locality}, which enables indirect reciprocity between peers in localized neighborhoods. Peers acknowledge each other's trustworthiness through the mechanism of acknowledgment of performed work. 
This paradigm allows the design to be Sybil-tolerant. 
Maximum trust in Sybil nodes is capped by the value of utility they can generate through a colluding bridge node that connects honest clients with Sybils. In any other scenario, Sybils must first generate utility before they are considered trustworthy, thereby negating the very conditions that enable the Sybil attack to be executed.

Our approach works through four mechanisms: selecting reputable peers, scoring peers based on interactions, gossiping said scores through cryptographically verifiable certificates, and ranking nodes based on these certificates and the MeritRank algorithm~\cite{nasrulin2022meritrank}. Together, they generate a level playing field, in which even unranked peers have the ability to rise to the level of the most reputable peers through our slot-based approach, which is adjusted in a round-based fashion.

Our experiments on the Bitcoin blockchain and the Tribler~\cite{tribler} network demonstrate how our accountability mechanism elegantly forms clusters of connections with reputable peers, which provide utility to the overall network while isolating selfish and malicious nodes. Our results show low time to convergence even with half of the selfish peers. While introducing negligible overhead on bandwidth and memory. Furthermore, our approach is generic enough to allow for applications in any type of peer-to-peer network.

\bibliographystyle{IEEEtran}
\bibliography{references}

\end{document}